\documentclass[11pt]{article}
\begin{document}
\begin{center}
\LARGE\bf Once Again On the Klein Paradox \vskip15mm
\large N.Kevlishvili$^1$, A. Khelashvili $^{1,2}$, T. Nadareishvili$^{1,2}$\\
\vspace{1cm} \small\sl
$^1$Department of Theoretical Physics,Tbilisi State University, Tbilisi, Georgia\\
$^2$High Energy Physics Institute, Tbilisi, Georgia\

\end {center}

\vspace{1cm}

\begingroup \addtolength{\leftskip}{1cm} \addtolength{\rightskip}{1cm}
\begin {center}
{\subsection*{Abstract}}
\end {center}
After the short survey of the Klein Paradox in 3-dimensional
relativistic equations, we present a detailed consideration of
Dirac modified equation, which follows by one particle infinite
overweighting in Salpeter Equation. It is shown, that the
separation of angular variables and reduction to radial equation
is possible by using standard methods in momentum space.\\
The kernel of the obtained radial equation differs from that of
spinless Salpeter equation in bounded regular factor. That is why
the equation has solutions of confined type  for infinitely
increasing potential.
\endgroup
 \vspace{1cm}

\section{Introduction}
It is well known, that the Dirac equation in central symmetric
field, which is a zero component of the Lorentz four-vector and is
infinitely increasing, has a pathological property - the leakage
through the infinite wall occurs or equivalently, the wave
function has an oscillating (i.e. non normalizable) asymptotic
behaviour. This unusual property is known as the Klein Paradox
\cite{Bjorken}.The same drawback has two-particle Dirac (Breit)
equation \cite{Krolikowski, Khelashvili1}. In case of quarks and
antiquarks this problem is usually avoided by introducing
additional Lorentz-scalar interaction \cite{Rein,Magyary}. Besides
this, particular importance is given to the equal mixture of
scalar and vector potentials, which excludes the spin-orbital
coupling in Dirac equation \cite{Khelashvili1,Magyary} and avoids
the Klein Paradox in Breit equation as well, \cite{Krolikowski,
Khelashvili1, Khelashvili2}.

It seems quite natural to ask for relativistic covariant QFT
equations for bound states such as Bethe-Salpeter (BS) equation.
There is an opinion that only vector exchange interaction is not
desirable here due to the Klein paradox. But as far as we know
this fact has not been proved regularly for BS equation yet. At
the same time, it is clear that if the Klein Paradox exists here,
the situation would be rather vague because the fundamental
particles that transfer interaction between quarks are vectorial
gluons and they must provide confinement of quarks into hadrons.
Otherwise all attempts to obtain singular (f.e. $q^{-4}$)
behaviour of gluon propagator \cite{AG} in the infrared area would
be useless.

In recent years a lot of articles have been dedicated to
3-dimensional relativistic equations which follow from BS equation
by using various reduction methods. These equations are
interesting because they give the possibility to study quarkonium
spectra for different potentials.

Despite the fact that  3-dimensional relativistic equations have a
long history of study in recent years much attention has been paid
to the formulations in which by infinite overwheighting  of one of
the constituent particles the problem is reduced to the Dirac
equation for light particle in external field
\cite{Kopaleishvili}.

This paradigm is not clear to us, because in our opinion, the
information on second quantization could not be lost completely
whereas one of the particles becomes infinitely heavier. Really
when we use 3-dimensional (for example, instantaneous) kernel in
BS equation the latter is reduced to Salpeter equation which
differs from two-particle Dirac (Breit) equation by projection
operators in interaction term only on positive and negative
frequencies, but do not contain their interference. These
operators are the only remnants of secondary quantization and they
cannot be neglected.

However in the 1960-s the difference caused by projection operator
for mesons in quark model was found. For instance, P. Horwitz
\cite{Horwitz} in case of square wall potential has shown that if
one keeps only $v^2/c^2$ terms in projection operators then some
unpleasant features of Breit equation disappear, as a consequence
of correct projecting, which excludes transitions from positive to
negative frequency states. That is why the effective potential
produced in this problem has correct sign and is free from
unphysical singularities.

Although this result had been derived only in $v^2/c^2$ order, it
was, by our knowledge, the first indication that there is a puzzle
practically equivalent to the Klein paradox.

Later the whole attention was carried on Salpeter equation. It was
supposed \cite{Khelashvili1,Khelashvili2} that QFT equations must
be free from the Klein paradox. There are remarkable works
\cite{Krolikowski1,Turski} dedicated to the study of Salpeter
equation from this point of view. However final conclusions are
made based on the above mentioned modified Dirac equation. This
subject was considered also in many other publications
\cite{sucher,AOVW,Olsson}.

For example, the author of ref. \cite{Krolikowski1} studied the
following Salpeter equation

\begin{equation}\label{F}
[E-(\vec{\alpha}^{(1)}\vec{p}+\beta^{(1)}m_1)-
(-\vec{\alpha}^{(2)}\vec{p}+\beta^{(2)}m_2)-
\Pi(\vec{p})V(r)]\psi\vec{(r)}=0\;,
\end{equation}
where $\Pi(\vec{p})$ is the well known projection operator

\begin{equation}
\Pi(\vec{p})=\Lambda^{(1)}_+(\vec{p})\Lambda^{(2)}_+
(\vec{-p})-\Lambda^{(1)}_-(\vec{p})\Lambda^{(2)}_-(\vec{-p})\;,
\end{equation}
and

\[\Lambda^{(i)}_\pm(\vec{p})=\frac{\omega_p\pm(\vec{\alpha}^{(i)}\vec{p}
+\beta^{(i)}m_i)}{2\omega_p},{~~~~~}
\omega_p=\sqrt{\vec{p}^2+m^2}\]
are corresponding projection
operators for free spinors. They represent nonlocal integral
operators in coordinate space.

In the limit when $m_1=m_2=m\rightarrow\infty$ the kernel of this
integral representation (Bessel $K_0(m|\vec{r}-\vec{r'}|)$
function) transforms into local $\delta(\vec{r}-\vec{r'})$
function, then it is not difficult to determine the asymptotic
behavior of wave function when $r\rightarrow\infty$. It falls
exponentially for infinitely increasing potentials, V(r).

In the other work \cite{Turski} the case $m_2\rightarrow\infty$ is
considered, when equation (\ref{F}) is reduced to modified Dirac
equation, describing a lighter particle's motion in "projected"
potential:

\begin{equation}\label{Th}
[\vec{\alpha}\vec{p}+{\beta}m+\Lambda_+{(\vec{p})}V(r)]\psi({\vec{r}})=E\psi({\vec{r}})\;
\end{equation}

By means of the above mentioned integral representation and some
additional asymptotic restrictions on the wave function and
potential the author shows, that modified Dirac equation is free
from the Klein paradox.

In our opinion all these statements are somewhat unsatisfactory,
because they are based on some approximations and/or simplified
assumptions. One thing is clear from these works - if modified
Dirac equation is free from the Klein paradox, the same will be
true for Salpeter equation.

Further we will consider our problem in momentum space and show
that modified Dirac equation (\ref{Th}) has only discrete spectrum
provided Schrodinger equation had such spectrum for the same
potentials V(r). In our opinion, this statement is equivalent to
the absence of the Klein paradox.

\section{Modified Dirac Equation in Foldy-Wouthuyson Representation}\label{chapf}

Consequently, let us study modified Dirac  equation (\ref{Th}). If
we successively act on this equation with projecting operators
$\Lambda_\pm(p)$, we get the following equation

\begin{equation}\label{For}
[E-(\vec{\alpha}\vec{p}+{\beta}m)]\Lambda_+{(\vec{p})}\psi({\vec{r}})=\Lambda_+{(\vec{p})}V\psi({\vec{r}});
\end{equation}
and the additional constraint

\begin{equation}
\psi_-\equiv\Lambda_-\psi=0
\end{equation}
Hence equation (\ref{For}) looks like

\begin{equation}
[E-(\vec{\alpha}\vec{p}+{\beta}m)]\Lambda_+\psi=\Lambda_+V\Lambda_+\psi;
\end{equation}

Thus, the task is reduced to the problem on eigenfunctions and
eigenvalues of the following hermitian Hamiltonian

\begin{equation}\label{Se}
H=(\vec{\alpha}\vec{p}+{\beta}m)+\Lambda_+V\Lambda_+;
\end{equation}

It is convenient to use the Foldy-Wouthuysen (FW) transformation
\cite{FW}

\begin{equation}
e^{iS}(\vec{\alpha}\vec{p}+{\beta}m))e^{-iS}=\beta\omega_p,{~~~~~}
\omega_p=\sqrt{\vec{p}^2+m^2}
\end{equation}

It is clear, that

\begin{eqnarray}
e^{iS}\Lambda_+(\vec{p})=\sqrt{\frac{2\omega_p}{{\omega_p+m}}}
\frac{1}{2}(1+\beta)\Lambda_+(\vec{p})\nonumber\\
\Lambda_+(\vec{p})e^{-iS}=\sqrt{\frac{2\omega_p}{{\omega_p+m}}}
\Lambda_+(\vec{p})\frac{1}{2}(1+\beta)
\end{eqnarray}

For the transformed $\psi_{FW}=e^{iS}\psi$ wave function we get
the following equation

\begin{eqnarray}\label{Ten}
\lefteqn{(E-\beta\omega_p)\psi_{FW}=}\\
&=\sqrt{\frac{2\omega_p}{{\omega_p+m}}}\frac{1}{2}(1+\beta)
\Lambda_+(\vec{p})\int{d^3kV(\vec{p}-\vec{k})\Lambda_+(\vec{k})\frac{1}{2}(1+\beta)}
\sqrt{\frac{2\omega_k}{{\omega_k+m}}}\psi_{FW}(k)\nonumber\
\end{eqnarray}
Now it is natural to use the 2-component representation

\begin{equation}
\psi_{FW}=\left(\begin{array}{c}\varphi\\\chi\end{array}\right)\
\end{equation}
Then equation (\ref{Ten}) transforms to the following system:

\begin{eqnarray}\label{Tw}
\lefteqn{(E-\omega_p)\varphi(\vec{p})=}\nonumber\\
&=\sqrt{\frac{2\omega_p}{{\omega_p+m}}}\frac{1}{2}(1+\beta)
\Lambda_+(\vec{p})\int{d^3kV(\vec{p}-\vec{k})\Lambda_+(\vec{k})
\sqrt{\frac{2\omega_k}{{\omega_k+m}}}\varphi(\vec{p})};\\
\lefteqn{(E+\omega_p)\chi(\vec{p})=0}\nonumber\
\end{eqnarray}

As far as $E\neq-\omega_P$, the second equation has only trivial
solution $\chi=0$.

Let us calculate the matrix structure more explicitly on the
right-hand side of equation (\ref{Tw}). We get

\begin{eqnarray}\label{TT}
\lefteqn{(E-\omega_p)\varphi(\vec{p})=}\\
&=\sqrt{\frac{2\omega_p}{{\omega_p+m}}}\int{d^3k[\frac{\omega_p+m}{2\omega_p}
V(\vec{p}-\vec{k})\frac{\omega_k+m}{2\omega_k}+
\frac{\vec{\sigma}\vec{p}}{2\omega_p}V(\vec{p}-\vec{k})
\frac{\vec{\sigma}\vec{k}}{2\omega_k}]\sqrt{\frac{2\omega_k}
{{\omega_k+m}}}\varphi(\vec{p})}\nonumber\
\end{eqnarray}

Let us make a remark that the same problem with the Hamiltonian
(\ref{Se}) has been considered earlier by J.Sucher \cite{sucher} .
The author using transformations close to the above mentioned had
obtain the equation similar to (\ref{TT}) except the difference in
some kinematical factors. To our opinion after all author made
somewhat premature conclusion: "Since these corrections do not
dominate the effective interaction, one expects that there are
normalizable solutions both in the scalar case and in the vector
case".

Below we bring alternative (and practically equivalent) proof for
the fourth component of the vector case. As it was mentioned in
the introduction, just vector case is the principle one. In
addition we remark that due to the $\frac{1+\beta}{2}$ factor
scalar potential is actually presented in our study.

For this purpose we use the separation of angular variables and
correspondingly, one dimensional radial equation in momentum
representation, in which one additional nontrivial property of
this equation will appear.

\section{The radial form of modified Dirac equation}\label{chapf}

An angular analysis can be easily performed in standard way. For
this purpose we should choose the basis of spherical spinors
\cite{AB}

\begin{equation}
\varphi(\vec{p})=f(p)\Omega_{jlM}(\vec{n_p}),
~~~\vec{n_p}=\frac{\vec{p}}{p}
\end{equation}
where $\Omega_{jlM}$ functions satisfy the following equations

\begin{equation}\label{line}
(\vec{\sigma}\vec{n_p})\Omega_{jlM}(\vec{n_p})=-\Omega_{jl'M}(\vec{n_p}),
~~~l+l'=2j
\end{equation}

These functions are expressed in explicit form by spherical
harmonics and corresponding Klebsh-Gordan coefficients:

\begin{equation}
\Omega_{jlM}(\vec{n})=\left(\begin{array}{cc}C^{jM}_{l,M-1/2,1/2,1/2}&Y_{l,M-1/2}(\vec{n})\\
C^{jM}_{l,M+1/2,1/2,-1/2}&Y_{l,M+1/2}(\vec{n})\end{array}\right)
\end{equation}

Then we present $V(\vec{p}-\vec{k})$ as series of spherical
harmonics

\begin{equation}
V(\vec{p}-\vec{k})=\sum^\infty_{l=0}\sum^l_{m'=-l}v_l(p,k)Y_{lm'}(\vec{n_p})
Y^*_{lm'}(\vec{n_k})
\end{equation}
and insert everything into the equation (\ref{TT}). Accounting
orthogonality and completeness properties it is easy to verify
that angular dependence is avoided and we get the following radial
equation in momentum space

\begin{equation}\label{Et}
(E-\omega_p)f(p)=\int^\infty_0k^2dkv_l(p,k)\chi(p,k)f(k)
\end{equation}
where $\chi(p,k)$ is an additional factor resulting from
projection operator and FW transformation,

\begin{equation}
\chi(p,k)=\sqrt{\frac{\omega_p+m}{2\omega_p}}[1+\frac{pk}{(\omega_p+m)(\omega_k+m)}]
\sqrt{\frac{\omega_k+m}{2\omega_k}}
\end{equation}

Equation (\ref{Et}) is the main result of the present paper.

It seems that there is some unexpectedness in this equation.
Particularly the fact that in the equation (\ref{TT}) matrix
structure

\[\frac{\vec{\sigma}\vec{p}}{2\omega_p}V(\vec{p}-\vec{k})\frac{\vec{\sigma}
\vec{k}}{2\omega_k}\] has been reduced completely and only the
dependence on orbital-angular momentum remains. The non-total
diagonalization was the more expectative outcome. But it seems
that the projection operators $\Lambda_+(p)$ tear the spin-orbital
coupling and restore the symmetry under usual 3-dimentional
rotations.

The argument for this statement may be the observation derived in
the recent years in Dirac equation with $V_v=\pm V_s$ potential
(or $\frac{1\pm\beta}{2}$ projectors), the so called pseudospin
symmetry \cite{GL,RGG}. Anyway, the symmetries of the Hamiltonian
under consideration need more detailed investigation.

\section{Properties of spectrum of radial equation}\label{chapf}

The type of spectrum of radial (integral) equation (\ref{Et})
depends on the properties of its kernel, which now contains
FW-factor $\chi(p,k)$ together with radial component of potential.
Just this factor expresses nonlocality of effective potential.

First of all let us list several properties of $\chi(p,k)$ factor:

1) It represents the sum of two terms factorized in p and k
variables. This property could be important during the study of
various separable potentials.

2) When k=p (on the energy-shell in elastic scattering problem),
$\chi=1$.

3) It is positively-definite and bounded when
$p,k\rightarrow\infty$.

4) It has no singularity for physical values of variables.

5) To reach the nonrelativistic limit in equation(\ref{TT}) it is
enough to expand it into the powers of small $p^2$ and $k^2$. This
operation does not imply smallness of potential V in contrast to
the case of pure Dirac equation.

Based on these properties we can answer our main (principal)
question if the Klein paradox takes place or not.

It is clear, that integral equation (\ref{Et}) without $\chi(p,k)$
is a radial form of Salpeter spinless equation in momentum space.

As it is known \cite{NDD} Salpeter spinless equation for
infinitely increasing potentials has confinement like solutions,
i.e. only discrete spectrum. Due to the above mentioned properties
of $\chi(p,k)$ the kernel of our equation is a product of
completely continuous type kernel on bounded nonsingular
functions. Therefore according to the well-known theorem
\cite{Lovelace,DS} the obtained kernel will be completely
continuous type, as well. We can conclude that the spectrum of our
equation (\ref{Tw}) will be only discrete for such potentials if
for those the Schrodinger (or spinless Salpeter) equation had the
discrete spectrum (see, e.g. Theorem 7.8.3 in book \cite{see}).

This statement, in our opinion, is equivalent to the absence of
Klein paradox in modified Dirac equation, because it means that
the primary equation has only normalizable solutions.


\end {document}